\documentclass[12pt]{article}

\usepackage{graphicx}
\begin{document}

\begin{center}
{\bf  New entropy, thermodynamics of apparent horizon and cosmology} \\
\vspace{5mm} S. I. Kruglov
\footnote{E-mail: kruglov@rogers.com}
\underline{}
\vspace{3mm}

\textit{Department of Physics, University of Toronto, \\60 St. Georges St.,
Toronto, ON M5S 1A7, Canada\\
Canadian Quantum Research Center, \\
204-3002 32 Ave., Vernon, BC V1T 2L7, Canada} \\
\end{center}

\begin{abstract}
Here, we consider new nonadditive entropy of the apparent horizon $S_K=S_{BH}/(1+\gamma S_{BH})$ with $S_{BH}$ being the Bekenstein--Hawking entropy. This is an alternative of the R\'{e}nyi and Tsallis entropies, that allows us, by utilising the holographic principle, to develop
a new model of entropic (holographic) dark energy. When $\gamma\rightarrow 0$ our entropy becomes the Bekenstein--Hawking entropy $S_{BH}$. The generalised Friedmann equations for Friedmann--Lema\^{i}tre--Robertson--Walker (FLRW) spacetime for the barotropic matter fluid with $p=w\rho$ were obtained. We compute the dark energy pressure $p_D$, density of energy $\rho_D$, the normalized density parameters $\Omega_D$, $\Omega_{m}$ and the deceleration parameter $q$ of the universe corresponding to our model. From the second modified Friedmann equation a dynamical cosmological constant was obtained. We show that at some model parameters $w$ and $\gamma$ we obtain $\Omega_{m0}\approx 0.315$ and $q_0\approx -0.535$ which are in agreement with the Planck data. It was shown that the model under consideration possesses the phantom divide for the EoS of dark energy. Thus, our model, by virtue of the holographic principle, can describe the universe inflation or the late time of the universe acceleration but simultaneous description of inflation and the late time of the universe acceleration in the current model is questionable. It is shown that entropic cosmology with our entropy proposed is equivalent to cosmology based on the teleparallel gravity with the function $F(T)$ obtained. The holographic dark energy model with the generalised entropy of the apparent horizon can be of interest for new cosmology.
\end{abstract}

\section{Introduction}

To describe the current universe acceleration, one can introduce in the Einstein equation the cosmological constant. It plays the role of dark energy leading to standard cosmology with its large scale homogeneity and isotropy. It is worth noting that our universe is homogeneous and isotropic on scales larger than $100$ Mpc \cite{Mukhanov}. According to the Big Bang scenario there was homogeneous and isotropic distribution of a matter at high temperature and density about $15$ billion years ago. Then the universe has been expanding and cooling.
The Friedmann equations with the positive cosmological constant lead to de Sitter spacetime which explains the current universe acceleration (due to the dark energy). Another way to describe acceleration of universe is to explore thermodynamics of apparent horizon \cite{Akbar,Cai,Cai0,Paranjape,Sheykhi,Cai1,Wang,Jamil,Gim,Fan,Agostino,Sanchez} because there is a correspondence between gravity and thermodynamics. This is based on the fact that as was shown in Refs. \cite{Bekenstein,Hawking}, in black holes the entropy is connected with the horizon area and the temperature is linked with the surface gravity \cite{Jacobson,Padmanabhan,Padmanabhan1,Hayward}.
Also, Friedmann's equations may be obtained with the help of the first law of apparent horizon thermodynamics. As a result, the apparent horizon of the FLRW spacetime is represented as a thermodynamic system and it is a causal boundary \cite{Hayward,Hayward1,Bak}, and on this boundary the thermodynamics laws are satisfied \cite{Akbar,Cai0}. In spatially flat FLRW universe the apparent horizon is equal to the Hubble horizon. Different entropies \cite{Tsallis,Barrow,Renyi,Kaniadakis,Masi,Czinner,Kruglov,Kruglov1} were introduced, due to the long-range nature of gravity, that lead to the generalized Friedmann equations. Thus, Tsallis entropy represents effective entropy for non-extensive thermodynamic systems with long range interactions. The Barrow entropy is similar to the Tsallis entropy and has an application in quantum gravity. The R\'{e}nyi entropy is connected with the amount of information of the system. The Kaniadakis entropy appears in the consideration of relativistic statistical systems.
The Sharma--Mittal entropy \cite{Masi} is two-parameter entropy which is a combination of the R\'{e}nyi and Tsallis entropies. The entropy of Loop Quantum Gravity \cite{Czinner} was used in black hole physics and cosmology. The entropies proposed in Ref. \cite{Kruglov,Kruglov1} are used in entropic cosmology and does not possesses a singularity at the Hubble rate $H=0$.
Previous studies of entropies and holographic dark energy models were considered in Refs. \cite{Jahromi,Ren,Mejrhit,Majhi,Sekhmani,Gashti,Sadeghi,Pourhassan,Sakalli,Sakalli_1,Sakalli_2,Sakalli_3,Khodam,Muhammad,Ahmad}. Motivated by success of this approach, it is of interest to consider other entropies to describe accelerating universe. Because entropies are sources of holographic energy densities they can describe the dark energy of the universe \cite{Pavon,Landim}. In this paper, we consider new apparent horizon entropy $S_K=S_{BH}/(1+\gamma S_{BH})$ where $S_{BH}$ is the Bekenstein--Hawking entropy and it is a non-extensive entropy measure. When the Bekenstein--Hawking entropy becomes zero our entropy $S_K$ also vanishes and $S_K$ is the monotonically increasing function of $S_{BH}$ and is positive. As $\gamma\rightarrow 0$ we come to the Bekenstein--Hawking entropy. Entropy and Friedmann equations in the current paper and in Refs. \cite{Kruglov,Kruglov1} are different. Thus, the entropy of Ref. \cite{Kruglov} is $S_h=\arctan(\beta S_{BH})/\beta$ and entropy of Ref. \cite{Kruglov1} is $S_h= S_{BH}/(1+ \lambda S_{BH}^2)$. Therefore, the evolution of universe is described in the different way here. In addition, the asymptotic of the entropy of Ref. \cite{Kruglov} at $R_h\rightarrow\infty$ is $\pi/(2\beta)$ and the asymptotic of the entropy of Ref. \cite{Kruglov1} at $R_h\rightarrow\infty$ is $0$. The asymptotic of entropy here at $R_h\rightarrow\infty$ is $1/\gamma$ and is monotonically increasing function as should be according to thermodynamics but the entropy of Ref. \cite{Kruglov1} is not.

Here, we consider the case with equation of state (EoS) for barotropic perfect fluid, $p=w\rho$. We will obtain modified Friedmann's equations with dynamical cosmological constant that leads to dark energy and the early and late time universe acceleration. The model under consideration is a modification of the model \cite{Kruglov1} and uses dimensionless variables that are convenient for the analyse.
Our new holographic model with the non-additive entropy shows the possibility of the universe to accelerate in accordance with observations.

We assume units with $\hbar=c=k_B=1$.

\section{New entropy}

We start with new entropy
\begin{equation}
S_K=-\sum_{i=1}^W\frac{p_i\ln {p_i}}{1-\gamma \ln {p_i}},
\label{1}
\end{equation}
where $W$ is a number of states. Each state possesses a probability $p_i$ with the probability distribution $\{p_i\}$ and $\gamma$ is a free parameter. In Eq. (1) the summation is performed over all possible system microstates. When $\gamma=0$ entropy (1) is converted into the Gibbs entropy
\begin{equation}
S_G=-\sum_{i=1}^Wp_i\ln (p_i).
\label{2}
\end{equation}
If one assumes that each microstate is  populated with equal probability, then $1/p_i=W$ ($i=1,2,...,W$) and Eq. (2) is converted into the Boltzmann entropy $S_B=\ln(W)$. Making use of $1/p_i=W$, we obtain from Eq. (1) equation as follows:
\begin{equation}
S_K=\frac{\ln(W)}{1+\gamma\ln (W)}.
\label{3}
\end{equation}
The Bekenstein--Hawking entropy is given by $S_{BH}=\ln(W)$. Then from Eq. (3) one finds
\begin{equation}
S_K=\frac{S_{BH}}{1+\gamma S_{BH}}.
\label{4}
\end{equation}
Making use of Eq. (4), at $\gamma=0$, we obtain the Bekenstein--Hawking entropy $S_{BH}=A/(4G)$, where $A=4\pi R_h^2$ is the area of the horizon. If $B$ and $C$ are two probabilistically independent systems, we have $p^{B+C}_{ij}=p^B_ip^C_j$ and one finds nonadditive entropy, $S_K(B+C)\neq S_K(B)+S_K(C)$. It is worth noting that the  Bekenstein--Hawking entropy $S_{BH}$ as well as entropies \cite{Tsallis,Barrow,Renyi,Kaniadakis} possess a singularity when the Hubble parameter vanishes, $H=0$, but  our entropy becomes $S_K=1/\gamma$ so that the singularity is absent.

\section{Apparent horizon thermodynamics}

Let us consider the FLRW spatially flat universe which is described by the metric
\begin{equation}
ds^2=-dt^2+a(t)^2(dr^2+r^2d\Omega_2^2).
\label{5}
\end{equation}
Here, $a(t)$ is a scale factor and $d\Omega_2^2$ is the line element of an 2-dimensional unit sphere. The radius of the apparent horizon $R_h = a(t)r$, in  the FLRW universe, is given by
\begin{equation}
R_h=\frac{1}{H},
\label{6}
\end{equation}
where the Hubble parameter of the universe, which measures the expansion rate, is $H=\dot{a}(t)/a(t)$, with dot over $a(t)$ being the derivative with respect to the cosmological time $t$. The first law of apparent horizon thermodynamics is formulated as
\begin{equation}
dE=-T_hdS_h+WdV_h,
\label{7}
\end{equation}
where $W$ is the work density and $E$ is the total energy inside the space which is given by
\begin{equation}
E=\rho V_h=\frac{4\pi}{3}\rho R_h^3.
\label{8}
\end{equation}
The change of the energy inside the apparent horizon is $dQ=-dE$. In Eq. (8) $\rho$ means the energy density of a matter and the work density is \cite{Hayward,Hayward1,Bak}
\begin{equation}
W=-\frac{1}{2}\mbox{Tr}(T^{\mu\nu})=\frac{1}{2}(\rho-p),
\label{9}
\end{equation}
and $p$ is the matter pressure. In our case the horizon entropy is $S_h=S_K$. The apparent horizon temperature is defined as
\begin{equation}
T_h=\frac{H}{2\pi}\left|1+\frac{\dot{H}}{2H^2}\right|.
\label{10}
\end{equation}
The sign of the factor $1+\dot{H}/(2H^2)$ was discussed in Ref. \cite{Kruglov}.
Making use of first law of apparent horizon thermodynamics (7), and by combining Eqs. (8) and (9) with (10), one finds
\begin{equation}
\frac{H}{2\pi}\left|1+\frac{\dot{H}}{2H^2}\right|dS_h=-\frac{4\pi}{3H^3}d\rho+\frac{2\pi(\rho+p)}{H^4}dH.
\label{11}
\end{equation}
With the help of the continuity equation
\begin{equation}
\dot{\rho}=-3H(\rho+p),
\label{12}
\end{equation}
and Eq. (11), we obtain
\begin{equation}
\frac{H}{2\pi}\left|1+\frac{\dot{H}}{2H^2}\right|\dot{S}_h=-\frac{4\pi\dot{\rho}}{3H^3}\left(1+\frac{\dot{H}}{2H^2}\right).
\label{13}
\end{equation}

\section{Generalized Friedmann's equations}

Utilizing Eqs. (12), (13) and assuming that $1+\dot{H}/(2H^2)>0$ we obtain
\begin{equation}
\frac{H}{2\pi}\dot{S}_h=\frac{4\pi (\rho+p)}{H^2}.
\label{14}
\end{equation}
By using our entropy function
\begin{equation}
S_h=S_K=\frac{S_{BH}}{1+\gamma S_{BH}},
\label{15}
\end{equation}
where $S_{BH}=\pi R^2_h/G=\pi/(GH^2)$, we obtain from Eq. (14) the generalized Friedmann equation
\begin{equation}
\frac{\dot{H}}{(1+\gamma\pi/(GH^2))^2}=-4\pi G(\rho+p).
\label{16}
\end{equation}
At $\gamma=0$ in Eq. (16), we come to the first Friedmann equation for spatial flat universe within Einstein's gravity. Making use of Eq. (12) and after  integration of Eq. (16), we find the second generalized Friedmann equation
\begin{equation}
H^2-\frac{b^2}{H^2+b}-2b\ln\left(\frac{H^2+b}{b}\right)=\frac{8\pi G}{3}\rho,
\label{17}
\end{equation}
where $b=\pi\gamma/G$ and we have used the integration constant $C=2b\ln(b)$.
At $\gamma=0$ ($b=0$) one obtains from Eq. (17) the Friedmann equation of general relativity. We can represent Eq. (17) as follows:
\begin{equation}
H^2=\frac{8\pi G}{3}\rho+\frac{\Lambda_{eff}}{3}
\label{18}
\end{equation}
with
\begin{equation}
\Lambda_{eff}=\frac{3b^2}{H^2+b}+6b\ln\left(\frac{H^2+b}{b}\right).
\label{19}
\end{equation}
The $\Lambda_{eff}$ represents  the dynamical (effective) cosmological constant.
We depicted the dynamical cosmological constant $\Lambda_{eff}$ versus $H$ at $b=1,2,3$  in Fig. 1. We use here the Planckian units with $G=c=\hbar=k_B=1$ \cite{Mukhanov}.
\begin{figure}[h]
\includegraphics [height=2.0in,width=3.5in] {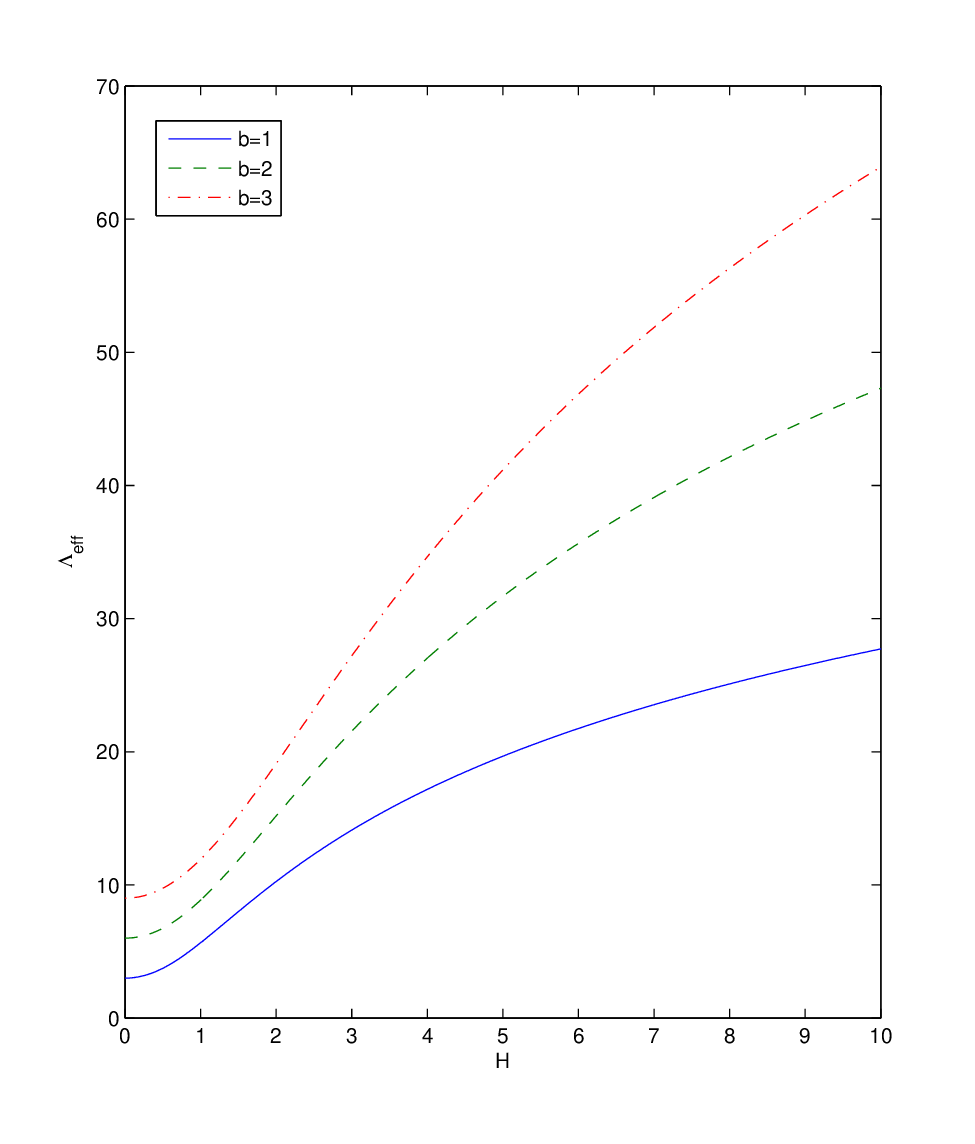}
\caption{\label{fig.1} The function $\Lambda_{eff}$ versus $H$ at $b=\pi \gamma/G=1,2,3$. Figure 1 shows that $\Lambda_{eff}$ increases as $b$ increases. At $H\rightarrow 0$ ($R_h\rightarrow\infty$) we have $\Lambda_{eff}\rightarrow 3b$.}
\end{figure}
According to Fig. 1 when $b$ increases, at fixed $H$, the dynamical cosmological constant $\Lambda_{eff}$ also increases and $\lim_{H\rightarrow 0} \Lambda_{eff}=3b$. Thus, at large apparent horizon radius $R_h$ (small $H$), we have the constant $\Lambda_{eff}$ explaining the current universe acceleration. By virtue of Eq. (18) one finds the density of dark energy
\begin{equation}
\rho_{D}=\frac{3b}{8\pi G}\left[\frac{b}{H^2+b}+2\ln\left(\frac{H^2+b}{b}\right)\right].
\label{20}
\end{equation}
Defining the normalized density parameters $\Omega_m=\rho/(3M_P^2 H^2)$ and $\Omega_D=\rho_D/(3M_P^2 H^2)$, where $M_P=1/\sqrt{8\pi G}$ is the reduced Planck mass, one obtains from Eqs. (17) and (20) that $\Omega_m+\Omega_D=1$.
From Eq. (17) we obtain the normalized density for the matter
\begin{equation}
\Omega_m=1-\frac{b^2}{H^2(H^2+b)}-2\frac{b}{H^2}\ln\left(\frac{H^2+b}{b}\right).
\label{21}
\end{equation}
By introducing new dimensionless variable $x=H^2/b$, Eq. (21) is rewritten as
\begin{equation}
\Omega_m=1-\frac{1}{x(1+x)}-\frac{2}{x}\ln(1+x).
\label{22}
\end{equation}
The plot of $\Omega_m$ versus $x$ is depicted in Fig. 2.
\begin{figure}[h]
\includegraphics [height=2.0in,width=3.5in] {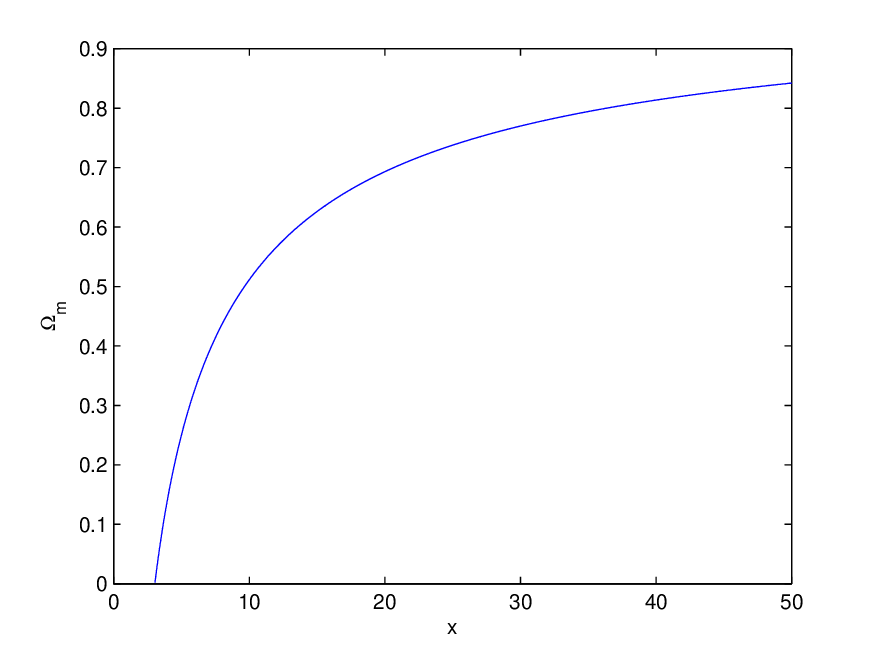}
\caption{\label{fig.2} The function $\Omega_m$ versus $x$. Figure 2 shows that $\Omega_m$ increases when $x$ increases.}
\end{figure}
In accordance with Fig. 2 as $x\rightarrow \infty$ ($H\rightarrow \infty$, $R_h\rightarrow 0$) one has $\Omega_m\rightarrow 1$. According to the Planck data \cite{Aghanim}, for the current era, $\Omega_{m0} \approx 0.315$. The solution to Eq. (22) at $\Omega_{m0}=0.315$ is given by $x\approx 5.819$. Then we obtain the entropy parameter $\gamma$,
\begin{equation}
\gamma=\frac{bG}{\pi}=\frac{GH_0^2}{5.819\pi}\approx 0.055~GH_0^2.
\label{23}
\end{equation}
The current value of the Hubble rate, according to the Planck data \cite{Aghanim}, is $H_0\approx 67.4$ km/s/Mpc.
We will imply here that there is no mutual interaction between various components of the cosmos, and from ordinary conservation law, one obtains
\begin{equation}
p_D=-\frac{\dot{\rho}_D}{3H}-\rho_D.
\label{24}
\end{equation}
By virtue of Eqs. (20) and (24) we find the pressure corresponding to the dark energy
\begin{equation}
p_D=-\frac{b(b+2H^2)\dot{H}}{4\pi G(b+H^2)^2}-
\frac{3b}{8\pi G}\left[\frac{b}{H^2+b}+2\ln\left(\frac{H^2+b}{b}\right)\right].
\label{25}
\end{equation}
Making use of Eqs. (16), (17) and (25) we obtain
\[
p_D=\frac{3b(b+2H^2)(1+w)}{8\pi GH^4}\biggl [H^2-\frac{b^2}{H^2+b}-
\]
\begin{equation}
2b\ln\left(\frac{H^2+b}{b}\right)\biggr ]-\frac{3b}{8\pi G}\left[\frac{b}{H^2+b}+2\ln\left(\frac{H^2+b}{b}\right)\right].
\label{26}
\end{equation}
With the help of Eqs. (20) and (26) one finds EoS for dark energy
\begin{equation}
w_D=\frac{p_D}{\rho_D}=\frac{(b+2H^2)(1+w)}{H^4}\left[\frac{H^2(H^2+b)}{b+2(H^2+b)\ln\left(\frac{H^2+b}{b}\right)}-b\right]-1.
\label{27}
\end{equation}
Making use of the variable $x=H^2/b$ we represent Eq. (27) as follows:
\begin{equation}
w_D=\frac{p_D}{\rho_D}=\frac{(1+2x)(1+w)}{x}\left[\frac{x+1}{1+2(x+1)\ln(x+1)}-\frac{1}{x}\right]-1.
\label{28}
\end{equation}
The plot of $w_D$ versus $x$ is depicted in Fig. 3.
\begin{figure}[h]
\includegraphics [height=2.0in,width=3.7in] {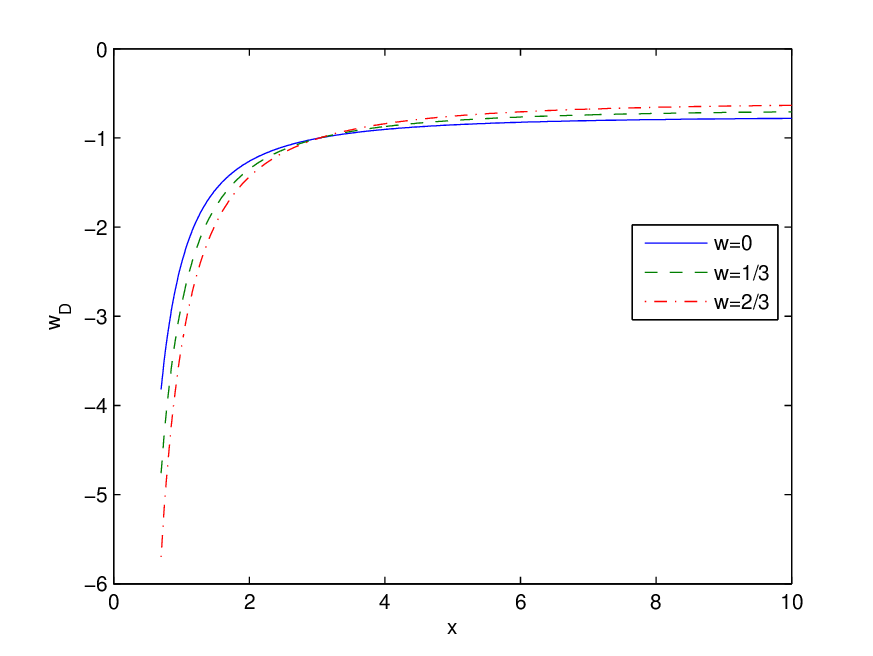}
\caption{\label{fig.3} The function $w_D$ versus $x=H^2/b$ at  $w=0, 1/3, 2/3$.
 At large $x$ the EoS parameter for dark energy $w_D$ approaches to $-1$, $\lim_{x\rightarrow\infty}w_D=-1$.}
\end{figure}
It follows from Eq. (28) that $\lim_{x\rightarrow\infty}w_D=-1$. Thus, dynamical cosmological constant leads to dark energy EoS $w_D=-1$ at large Hubble parameter $H$ (small $R_h$) which corresponds for the inflation era. Figure 3 shows that there is the phantom phase with $w_D<-1$. From Eq. (28) we obtain the equation for the phantom divide $w_D=-1$ as follows:
\begin{equation}
x(1+x)-2(x+1)\ln(x+1)-1=0.
\label{29}
\end{equation}
The solution to Eq. (29) is $x=H^2/b\approx 3.04$ (see Fig. 3). The phantom divide does not depend on the EoS of the matter $w$.
Thus, our model realizes the crossing of the phantom divide for the EoS of the dark energy.

The second law of thermodynamics requires that $\dot{S}_K\geq 0$. Then from Eq. (6) one finds $\dot{S}_{BH}/(1+\gamma S_{BH})^2\geq 0$ and this leads to $\dot{S}_{BH}=-2\pi\dot{H}/(GH^3)\geq 0$. As a result, we have the same requirement as for the Bekenstein--Hawking entropy. For positive Hubble parameter one obtains $\dot{H}\leq 0$. Then from Eq. (7) we find $\rho+p\geq 0$ and for $\rho>0$ the EoS parameter $w\geq -1$. It is convenient to use the redshift $z = a_0/a(t)-1$, where $a_0$ corresponds to a scale factor at the current time. With the aid of continuity equation (12) and EoS $p=w\rho$, one obtains the  matter density energy
\begin{equation}
\rho=\rho_0(1+z)^{3(1+w)},
\label{30}
\end{equation}
with $\rho_0$ being the density energy of matter at the present time. From Eqs. (17) and (30) we find equation as follows:
\begin{equation}
H^2-\frac{b^2}{H^2+b}-2b\ln\left(\frac{H^2+b}{b}\right)=\frac{8\pi G\rho_0}{3}(1+z)^{3(1+w)}.
\label{31}
\end{equation}
Equation (31) is the Friedmann equation that includes the dark energy density and determines $H(z)$.
According to Eq. (31) for the current era, $z=0$, the Friedmann equation does not depend on EoS of matter $w$. Therefore,  $\Omega_{m0}$ and $\Omega_{D0}$ are defined by Eq. (31) and do not depend on $w$. From Eq. (31) we obtain the redshift
\begin{equation}
z=\left[\frac{3}{8\pi\rho_0G}\left(H^2-\frac{b^2}{H^2+b}-2b\ln\left(\frac{H^2+b}{b}\right)\right)\right]^{1/(3(1+w))}-1.
\label{32}
\end{equation}
By introducing dimensionless parameters $\bar{H}=H/\sqrt{G\rho_0}$, $\bar{b}=b/(G\rho_0)$, Eq. (32) becomes
\begin{equation}
z=\left[\frac{3}{8\pi}\left(\bar{H}^2-\frac{\bar{b}^2}{\bar{H}^2+\bar{b}}-2\bar{b}\ln\left(\frac{\bar{H}^2+\bar{b}}{\bar{b}}\right)\right)\right]^{1/(3(1+w))}-1.
\label{33}
\end{equation}
The reduced Hubble parameter $\bar{H}$ versus redshift $z$ is plotted in Fig. 4.
\begin{figure}[h]
\includegraphics [height=2.0in,width=3.7in] {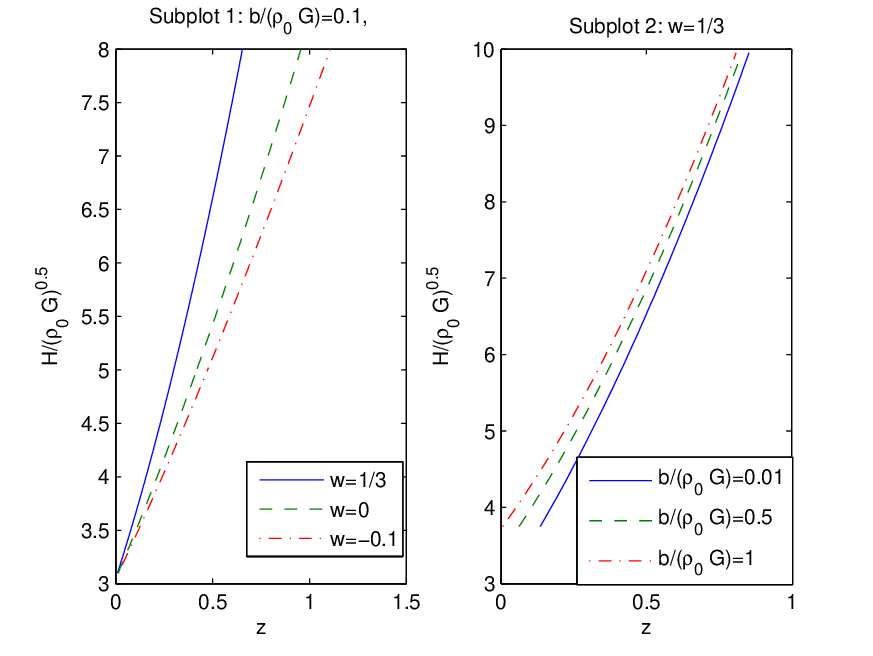}
\caption{\label{fig.4}\textbf{ Left panel}: The function $\bar{H}$ versus $z$ at $\bar{b}=0.1$, $w=1/3, 0, -0.1$.
According to Fig. 4, $\bar{H}$ increases as $z$ increases. At fixed $\bar{H}$, when EoS parameter for the matter $w$ increases the redshift $z$ decreases. \textbf{Right panel}: In accordance with figure if parameter $\bar{b}$ increases, at fixed $z$, the reduced Hubble parameter $\bar{H}$ also increases.}
\end{figure}
When redshift $z$ increases the reduced Hubble parameter also increases. According to Fig. 4 (Left panel) if parameter $w$ increases, at fixed $\bar{H}$, the redshift $z$ decreases. Figure 4 (Right panel) shows that when parameter $\bar{b}$ increases, at fixed z, the reduced Hubble parameter $\bar{H}$ also increases.

The deceleration parameter is given by
\begin{equation}
q=-\frac{\ddot{a}a}{\dot{a}^2}=-1-\frac{\dot{H}}{H^2}.
\label{34}
\end{equation}
If $q<0$ we have the acceleration phase of the universe and when $q>0$ the universe decelerates. Making use of Eqs. (16), (30) and (34) we find
\begin{equation}
q=\frac{4\pi G\rho_0(1+w)(H^2+b)^2}{H^6}\left(1+z\right)^{3(1+w)}-1.
\label{35}
\end{equation}
Equation (35) defines the dependence of the deceleration parameter $q$ on redshift $z$.
By virtue of Eqs. (31) and (35) we obtain the function of deceleration parameter $q$ on $H$
\begin{equation}
q=\frac{3(1+w)(H^2+b)^2}{2H^6}\left(H^2-\frac{b^2}{H^2+b}-2b\ln\left(\frac{H^2+b}{b}\right)\right)-1.
\label{36}
\end{equation}
Making use of dimensionless variable $x=H^2/b$, Eq. (36) becomes
\begin{equation}
q=\frac{3(1+w)(1+x)^2}{2x^2}\left(1-\frac{1}{x(1+x)}-\frac{2}{x}\ln(1+x)\right)-1.
\label{37}
\end{equation}
Taking into account $x=H_0^2/b\approx 5.819$ ($\gamma\approx H_0^2 G/(5.819\pi)$) which gives the normalized density for the matter field at the current time $\Omega_{m0}\approx 0.315$ and the deceleration parameter $q_0\approx -0.535$ \cite{Aghanim}, we obtain the solution to Eq. (37) for the EoS parameter of the matter $w\approx -0.2833$. Thus, the current values of $\Omega_{m0}$ and $q_0$ are in accordance with the Planck data at the parameter $b=H_0^2/5.819$, where $H_0\approx 67$~km/s/Mpc and the EoS parameter of the matter $w\approx -0.2833$. It should be noted that negative sign of EoS $w\approx -0.28$ indicates on the negative matter pressure  (not pressureless matter) so that dark energy component presents in $\Omega_{m0}$. This gives a limitation on Fig. 2 fit.

We depicted the deceleration parameter $q$  versus the $x$ in Fig. 5.
\begin{figure}[h]
\includegraphics [height=2.0in,width=3.7in] {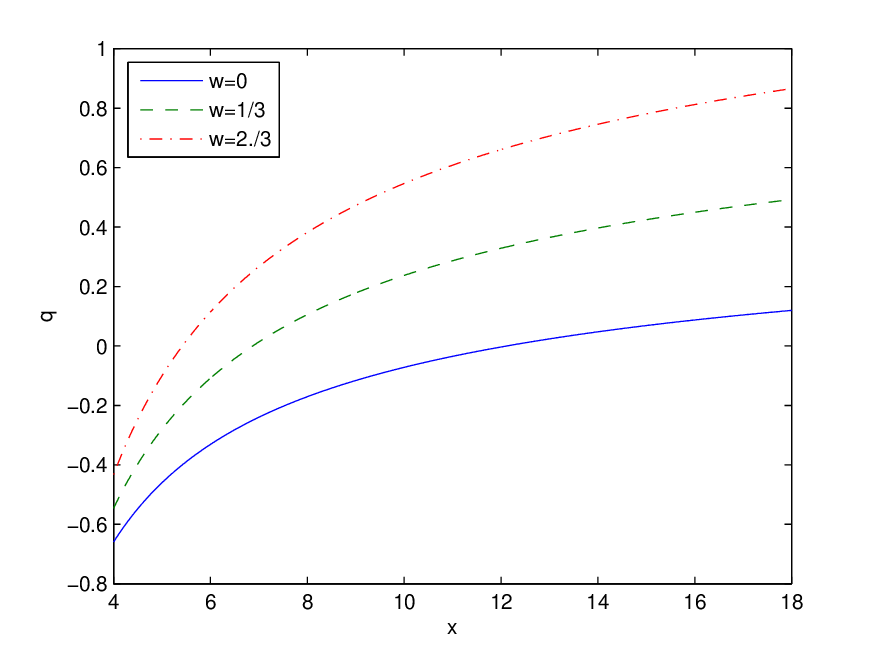}
\caption{\label{fig.5} The deceleration parameter $q$ versus $x$ at $w=0, 1/3, 2/3$. According to figure the deceleration parameter $q$ increases when the parameter $x$ increases.
When the EoS parameter of the matter $w$ increases, at fixed $x$, the deceleration parameter $q$ also increases. There are two phases: universe acceleration $q<0$ and deceleration $q>0$.}
\end{figure}
Figure 5 shows that there are two phases: the universe acceleration and deceleration.

Making use of Eq. (36) we find the asymptotic
\begin{equation}
\lim_{H\rightarrow \infty} q=\frac{3w+1}{2}.
\label{38}
\end{equation}
Equation (38) shows that the asymptotic of the deceleration parameter as $H\rightarrow \infty$ ($R_h\rightarrow 0$) does not depend on the entropy parameter $\gamma$. It follows from Eq. (38) that when $w>-1/3$ ($q>0$), at small $R_h$, we have the universe deceleration. Thus, the inflation of the universe (the universe acceleration at small $R_h$) takes place at $w<-1/3$.
By virtue of Eq. (37) at $q=0$ we obtain the equation for the transition phase
\begin{equation}
w=\frac{2x^3}{3(1+x)\left(x(1+x)-2(1+x)\ln(1+x)-1\right)}-1.
\label{39}
\end{equation}
We plotted the EoS parameter for the matter $w$ versus $x$ in Fig. 6.
\begin{figure}[h]
\includegraphics [height=2.0in,width=3.7in] {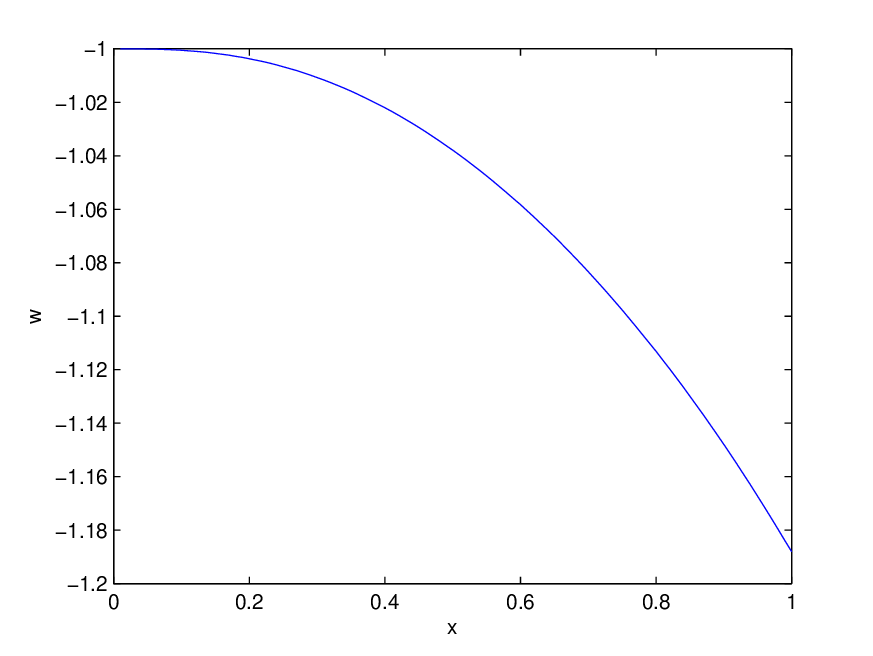}
\caption{\label{fig.6} The EoS parameter for the matter $w$ verses $x$ at $q=0$. When $x$ increases from $x=0$ ($H=0$), the EoS parameter $w$ decreases from $w=-1$ (de Sitter space) and becomes the phantom space. Thus, the transition from the acceleration phase at large $R_h$ (small $H$) into the deceleration phase is possible only for $w<-1$.}
\end{figure}
Figure 6 shows that when $x=H^2/b$ increases from $x=0$ the EoS parameter $w$ decreases from $w=-1$ (de Sitter space) and becomes the phantom space. As a result, the transition from the acceleration phase at large $R_h$ (small $H$) into the deceleration phase is possible only for phantom space $w<-1$. According to Fig. 5, for the bigger $x$ ($x>3$) the transition from $q<0$ to $q>0$ takes place at $w>0$.

\subsection{Inflation}

Inflation corresponds to the accelerated expansion of the universe and gravity possesses a repulsive force.
At the early stage of the universe, i.e. inflation, the matter is absent, $\rho=p=0$. The entropic energy (dark energy) density triggers the universe to accelerate. Then we obtain from Eq. (16) that $\dot{H}=0$, i.e. during inflation the Hubble parameter is a constant. As a result, the scale factor is given by $a(t)=a_0\exp(Ht)$ and corresponds to the de Sitter spacetime. It should be noted that de Sitter stage leads to the eternal inflation. The Hubble parameter at the early universe (inflation) is $H\approx10^{-3}M_{Pl}$ ($M_{Pl}=1/\sqrt{8\pi G}$). Then $GH^2\approx 4\cdot 10^{-8}$.
From Eq. (17), at $\rho=0$ and using the variable $x=H^2/b$  we obtain the equation as follows:
\begin{equation}
x-\frac{1}{x+1}-2\ln(x+1)=0,
\label{40}
\end{equation}
which is identical with the Eq. (29) for the phantom divide.
The non-trivial solution to Eq. (40) is given by $x=3.04002$ and we have the relation
\begin{equation}
H^2=3.04002 b.
\label{41}
\end{equation}
Knowing the Hubble parameter during universe inflation, $H\approx10^{-3}M_{Pl}$, we can fix parameter $b$ from Eq. (41) and the entropy parameter $\gamma=bG/\pi$. Thus, we obtain $\gamma\approx 4.2\times 10^{-9}$. In the scenario considered one can fix the constant $\gamma$ to describe inflation or late time universe acceleration. One can examine the generalized model when the constant $\gamma$ depends on the time in such a way that the model would describe both eras. We leave this scenario for further study.
It is worth noting that to have a viable model of inflation one needs a smooth graceful exit into the deceleration phase \cite{Mukhanov}.

\section{F(T)-gravity from generalized entropy}

In teleparallel gravity, a scalar torsion $T$ plays a role of a fundamental field similar to the curvature $R$ in Einstein's general relativity theory \cite{Hehl,Capozziello}. To describe the inflationary era and the current universe accelerating expansion, one introduces a Lagrangian in the form of $F(T)$ in analogy with $F(R)$-gravity. The teleparallel theory of gravity based on teleparallel geometry uses the Weitzenb\"{o}ck connection (not the Levi-Civita connection). In the $F(T)$ theory the field equations are the second order which is an advantage compared with $F(R)$ theory having the fourth order equations. The spacetime does not have the curvature and possesses the torsion.
The torsion scalar field $T$ is given by \cite{Weitzen,Maluf}
\begin{equation}
T=S_\rho^{~\mu\nu}T^\rho_{~\mu\nu},
\label{42}
\end{equation}
where tensors $S_\rho^{~\mu\nu}$ and $T^\rho_{~\mu\nu}$ are defined as follows:
\[
S_\rho^{~\mu\nu}=\frac{1}{2}\left(K_{~~\rho}^{\mu\nu}+\delta^\mu_\rho T_{~~\alpha}^{\alpha\nu}-\delta^\nu_\rho T_{~~\alpha}^{\alpha\mu}\right),
\]
\[
K^{\mu\nu}_{~~\rho}=-\frac{1}{2}\left(T^{\mu\nu}_{~~\rho}-T^{\nu\mu}_{~~\rho}-T^{~\mu\nu}_\rho\right),
\]
\begin{equation}
T^\rho_{~\mu\nu}=e^\rho_{i}\left(\partial_\mu e^i_{\nu}-\partial_\nu e^i_{\mu}\right),
\label{43}
\end{equation}
where $e^i_{\nu}$ ($i = 0, 1, 2, 3$) is  a vierbein field, so that the metric tensor is $g_{\mu\nu}=\eta_{ij}e^i_\mu e^j_\nu$ with $\eta_{ij}$ being the flat metric of the tangent spacetime. For FLRW metric (5), we have $e^i_\mu=\mbox{diag}(1,a,a,a)$. Then the torsion scalar field is given by $T=-6H^2$. From the action by variation with respect to $e^i_\mu$, where the Lagrangian is $F(T)$, one obtains \cite{Bengochea}
\begin{equation}
\frac{1}{6}\left[F(T)-2TF'(T)\right]|_{T=-6H^2}=\left(\frac{8\pi G}{3}\right)\rho.
\label{44}
\end{equation}
Making use of equations (17) and (44) we find
\begin{equation}
F(T)-2TF'(T)=-T-\frac{36b^2}{6b-T}-12b\ln\left(1-\frac{T}{6b}\right).
\label{45}
\end{equation}
By integration of Eq. (45) we obtain the function $F(T)$:
\begin{equation}
F(T)=T-3\sqrt{6bT}~\mbox{tanh}^{-1} \sqrt{\frac{T}{6b}}
-12b\ln\left(1-\frac{T}{6b}\right)+C\sqrt{T}-6b,
\label{46}
\end{equation}
where $C$ is the integration constant. It is worth noting that the term $C\sqrt{T}$ in Eq. (45) is the solution of the homogeneous equation $F(T)-2TF'(T)=0$ and should be eliminated, $C=0$. As $T=-6H^2<0$, we can use the formula $i~\mbox{tanh}^{-1}(ix)=-\arctan(x)$. Then Eq. (46) is converted into the real function (at $C=0$) as follows:
\begin{equation}
F(T)=T+3\sqrt{-6bT}~\arctan \sqrt{-\frac{T}{6b}}-12b\ln\left(1-\frac{T}{6b}\right)-6b.
\label{47}
\end{equation}
Some models of teleparallel gravity with different functions $F(T)$ were studied in \cite{Wu,Saridakis}.
Thus, the teleparallel gravity with the function (47) corresponds to entropic cosmology with entropy (15) under consideration.

\section{Conclusion}

Considering a novel entropy $S_K=S_{BH}/(1+\gamma S_{BH})$ we have a property similar to the Bekenstein--Hawking entropy $S_{BH}$.
By using our entropy and by employing the holographic principle, we have obtained a dark energy model.
The $S_K$ vanishes when the apparent horizon radius $R_h$ is zero and it increases monotonically when the apparent horizon radius $R_h$ increases. The barotropic perfect fluid and spatial flat FLRW universe were studied. We have obtained, from first law of apparent horizon thermodynamics, the modified Friedmann equations. The second Friedmann equation possesses an addition term corresponding to the density of dark energy which can be treated as a dynamical cosmological constant. Implying that there is no interaction between various components of cosmos, and the dark energy density and pressure $\rho_D$ and $p_D$ obeys ordinary conservation law, we computed $\rho_D$ and $p_D$. The EoS parameter for dark energy $w_D=p_D/\rho_D$ was calculated and it was shown that $\lim_{H\rightarrow\infty}w_D=-1$. Thus, at the small apparent horizon radius $R_h$ the de Sitter stage is realized which describes the inflation.
In the model under consideration the universe can have two phases, acceleration and deceleration, that is due to holographic dark energy. It should be noted that Barrow and Tsallis entropies also lead to cosmology which is due to Einstein's equations with the dynamical cosmological constant \cite{Gennaro}. We computed the deceleration parameter that shows the possibility, for some model parameters, to describe the acceleration at the current era. We show that at the entropy parameter $\gamma\approx 0.055 GH_0^2$ and $w=-0.283$ we have the deceleration parameter $q_0\approx -0.535$ and the normalized density parameter $\Omega_{m0}\approx 0.315$ which were observed at the current era \cite{Aghanim}. We showed that our model possesses the phantom divide for the EoS of dark energy.
Thus, our approach, based on new entropy and leading to modified Friedmann equations, can describe the universe inflation or the late time of universe acceleration. But the model does not describe inflation and the late time universe acceleration together.
It has been proven that entropic cosmology with our entropy proposed is equivalent to cosmology based on the teleparallel gravity with the function $F(T)$ (Eq. (47)).



\begin{thebibliography}{99}

\bibitem{Mukhanov} V. Mukhanov, Physical Foundations of Cosmology, Cambridge University Press (2005).
\bibitem{Akbar} M. Akbar and R. G. Cai, Thermodynamic Behavior of Friedmann Equation at Apparent Horizon of FRW Universe, Phys. Rev. D \textbf{75} (2007), 084003 [hep-th/0609128].
\bibitem{Cai} R. G. Cai and L. M. Cao, Unified First Law and Thermodynamics of Apparent Horizon in FRW Universe, Phys. Rev. D \textbf{75} (2007), 064008 [gr-qc/0611071].
\bibitem{Cai0} R. G. Cai and S. P. Kim, First Law of Thermodynamics and Friedmann Equations of Friedmann-Robertson-Walker Universe, JHEP \textbf{0502} (2005), 050 [arXiv:hep-th/0501055].
\bibitem{Paranjape} A. Paranjape, S. Sarkar and T. Padmanabhan, Thermodynamic route to Field equations in Lanczos-Lovelock Gravity, Phys. Rev. D \textbf{74} (2006), 104015 [hep-th/0607240].
\bibitem{Sheykhi} A. Sheykhi, B. Wang and R. G. Cai, Thermodynamical Properties of Apparent Horizon in Warped DGP Braneworld, Nucl. Phys. B \textbf{779} (2007), 1 [arXiv:hep-th/0701198].
\bibitem{Cai1} R. G. Cai and N. Ohta, Horizon Thermodynamics and Gravitational Field Equations in Horava-Lifshitz Gravity, Phys. Rev. D \textbf{81} (2010), 084061 [arXiv:0910.2307 [hep-th]].
\bibitem{Wang}S. Wang, Y. Wang and M. Li, Holographic Dark Energy, Phys. Rept. \textbf{696} (2017), 1 [arXiv:1612.00345 [astro-ph.CO]].
\bibitem{Jamil} M. Jamil, E. N. Saridakis and M. R. Setare, The generalized second law of thermodynamics in Horava-Lifshitz cosmology, JCAP \textbf{1011} (2010), 032 [arXiv:1003.0876 [hep-th]].
\bibitem{Gim} Y. Gim, W. Kim and S. H. Yi, The first law of thermodynamics in Lifshitz black holes revisited, JHEP \textbf{1407} (2014), 002.  [arXiv:1403.4704 [hep-th]].
\bibitem{Fan} Z. Y. Fan and H. Lu, Thermodynamical First Laws of Black Holes in Quadratically-Extended Gravities, Phys. Rev. D \textbf{91} (2015), 064009 [arXiv:1501.00006 [hep-th]].
\bibitem{Agostino} R. D’Agostino, Holographic dark energy from nonadditive entropy: cosmological perturbations and observational constraints, Phys. Rev. D \textbf{99} (2019), 103524 [arXiv:1903.03836 [gr-qc]].
\bibitem{Sanchez} L. M. Sanchez and H. Quevedo, Thermodynamics of the FLRW apparent horizon, Phys. Lett B \textbf{839} (2023), 137778 [arXiv:2208.05729 [gr-qc]].
\bibitem{Bekenstein} J. D. Bekenstein, Black Holes and Entropy, Phys. Rev. D \textbf{7} (1973), 2333-2346.
\bibitem{Hawking} S. W. Hawking, Particle creation by black holes, Commun. Math. Phys. \textbf{43} (1975), 199-220; Erratum: ibid. \textbf{46} (1976), 206.
\bibitem{Jacobson}T. Jacobson, Thermodynamics of Spacetime: The Einstein Equation of State, Phys. Rev. Lett. \textbf{75} (1995), 1260. [arXiv:gr-qc/9504004].
\bibitem{Padmanabhan} T. Padmanabhan, Gravity and the Thermodynamics of Horizons, Phys. Rept. \textbf{406} (2005), 49.
\bibitem{Padmanabhan1} T. Padmanabhan, Thermodynamical Aspects of Gravity: New insights, Rept. Prog. Phys. \textbf{73} (2010), 046901.  [arXiv:0911.5004 [gr-qc]].
\bibitem{Hayward} S. A. Hayward, Unified first law of black-hole dynamics and relativistic thermodynamics, Class. Quant. Grav. \textbf{15} (1998), 3147-3162 [arXiv:gr-qc/9710089 [gr-qc]].
\bibitem{Hayward1} S. A. Hayward, S. Mukohyana, M.C. Ashworth, Dynamic black-hole entropy, Phys. Lett. A \textbf{256} (1999), 347 [arXiv:gr-qc/9810006].
\bibitem{Bak} D. Bak, S. J. Rey, Cosmic holography, Class. Quant. Grav. \textbf{17} (2000), 83 [arXiv:hep-th/9902173].
\bibitem{Tsallis}C. Tsallis, Possible generalization of Boltzmann-Gibbs statistics, J. Stat. Phys., \textbf{52} (1-2) (1988), 479-487;  C. Tsallis, The Nonadditive Entropy $S_q$ and Its Applications in Physics and Elsewhere: Some Remarks, Entropy 13 (2011), 1765.
\bibitem{Barrow} J. D. Barrow, The Area of a Rough Black Hole, Phys. Lett. B \textbf{808} (2020), 135643 [arXiv:2004.09444 [gr-qc]].
\bibitem{Renyi} A. R\'{e}nyi, Proceedings of the Fourth Berkeley Symposium on Mathematics, Statistics and Probability, University of California Press (1960), 547-56.
\bibitem{Kaniadakis} G. Kaniadakis, Statistical mechanics in the context of special relativity II, Phys. Rev. E \textbf{72} (2005), 036108 [arXiv:cond-mat/0210467 [cond-mat.stat-mech]].
 \bibitem{Masi} Marco Masi, A step beyond Tsallis and R\'{e}nyi entropies, Phys. Lett. A \textbf{338} (2005), 217-224 [arXiv:cond-mat/0505107 [cond-mat.stat-mech].
\bibitem{Czinner} V. G. Czinner and H. Iguchi, R\'{e}nyi entropy and the thermodynamic stability of black holes, Phys. Lett. B \textbf{752} (2016), 306-310 [arXiv:1511.06963].
\bibitem{Kruglov} S. I. Kruglov, Cosmology Due to Thermodynamics of Apparent Horizon, Annalen der Phys. \textbf{534} (2025), e00204.
\bibitem{Kruglov1} S. I. Kruglov, Cosmology, new entropy and thermodynamics of apparent horizon, Chin. J. Phys. \textbf{98} (2025), 277-286.
\bibitem{Jahromi} A. Sayahian Jahromi, S. A. Moosavi, H. Moradpour, J. P. Morais Graca, I. P. Lobo, I. G. Salako and A. Jawad, Generalized entropy formalism and a new holographic dark energy model, Phys. Lett. B \textbf{780} (2018), 21-24 [arXiv:1802.07722].
 \bibitem{Ren} J. Ren, Analytic critical points of charged R\'{e}nyi entropies from hyperbolic black holes, JHEP \textbf{05} (2021), 080 [arXiv:2012.12892 [hep-th]].
\bibitem{Mejrhit} K. Mejrhit and S. E. Ennadifi,  Thermodynamics, stability and Hawking–Page transition of black holes from non-extensive statistical mechanics in quantum geometry, Phys. Lett. B \textbf{794} (2019), 45-49.
\bibitem{Majhi}A. Majhi, Non-extensive Statistical Mechanics and Black Hole Entropy From Quantum Geometry, Phys. Lett. B \textbf{775} (2017), 32-36  [arXiv:1703.09355 [gr-qc]].
\bibitem{Sekhmani}Yassine Sekhmani, et al., Exploring Tsallis thermodynamics for boundary conformal field theories in gauge/gravity duality, Chin. J. Phys. \textbf{92} (2024), 894–914.
\bibitem{Gashti} Saeed Noori Gashti, Behnam Pourhassan, \.{I}zzet Sakallı and Aram Bahroz Brzo, Thermodynamic Topology and Photon Spheres of Dirty Black Holes within Non-Extensive Entropy, Phys. Dark Univ. \textbf{47} (2025), 101833 [arXiv:2504.11939v1 [hep-th].
\bibitem{Sadeghi} J. Sadeghi, B. Pourhassan, and Z. Abbaspour Moghaddam, Interacting Entropy-Corrected Holographic Dark Energy and IR Cut-Off Length, Int. J. Theor. Phys. \textbf{53} (2014), 125–135 [arXiv:1306.2055 [gr-qc].
\bibitem{Pourhassan} B. Pourhassan, Alexander Bonilla, Mir Faizal, and Everton M. C. Abreu, Holographic Dark Energy from Fluid/Gravity Duality Constraint by Cosmological Observations, Phys. Dark Univ. \textbf{20} (2018), 41 [arXiv:1704.03281 [hep-th]].
\bibitem{Sakalli} I. Sakalli, M. Halilsoy, H. Pasaoglu, Entropy Conservation of Linear Dilaton Black Holes in Quantum Corrected Hawking Radiation, Int. J. Theor. Phys. \textbf{50} (2011), 3212.
\bibitem{Sakalli_1}Behnam Pourhassan, Hoda Farahani, Farideh Kazemian, Izzet Sakalli, Sudhaker Upadhyay, Dharm Veer Singh, Non-perturbative correction on the black hole geometry, Phys. Dark Univ. \textbf{44} (2024), 101444 .
\bibitem{Sakalli_2}Saeed Noori Gashti, Ankit Anand, Mohammad Ali S. Afshar, Mohammad Reza Alipour, Yassine Sekhmani, Behnam Pourhassan, Izzet Sakalli, Jafar Sadeghi, AdS-Schwarzschild-like black hole thermodynamics: Loop quantum gravity impact on topology and universality, Nucl. Phys. B \textbf{1021} (2025), 117188.
\bibitem{Sakalli_3} Erdem Sucu, Suat Dengiz, Izzet Sakalli, Quantum-corrected thermodynamics of Conformal Weyl Gravity black holes: GUP effects and phase transitions, Ann. Phys. (N.Y.) \textbf{490} (2026), 170494.
\bibitem{Khodam}A. Khodam-Mohammadi, M. Monshizadeh, Exploring modifications to FLRW cosmology with general entropy and thermodynamics: A new approach,  Phys. Lett. B \textbf{843} (2023), 138066.
\bibitem{Muhammad}Muhammad Naeem, Aysha Bibi, Correction to the Friedmann equation with Sharma–Mittal entropy: A new perspective on cosmology, Ann. Phys. (N.Y.) \textbf{462} (2024), 169618.
\bibitem{Ahmad} Ahmad Sheykhi, Thermodynamics of apparent horizon in mimetic cosmology, Int. J. Mod. Phys. D \textbf{28} (2019), 1950057.
\bibitem{Pavon}D. Pavon and W. Zimdahl, Holographic dark energy and cosmic coincidence, Phys. Lett. B \textbf{628} (2005), 206 [gr-qc/0505020].
\bibitem{Landim}R. C. G. Landim, Holographic dark energy from minimal supergravity, Int. J. Mod. Phys. D \textbf{25} (2016), 1650050 [arXiv:1508.07248[hep-th]].
\bibitem{Aghanim} N. Aghanim et al. [Planck], Cosmological parameters, Astron. Astrophys. \textbf{641} (2020), A6; Erratum: ibid, \textbf{652} (2021), C4 [arXiv:1807.06209 [astro-ph.CO]].
\bibitem{Gennaro} Sofia Di Gennaro, Hao Xu, Yen Chin Ong, How barrow entropy modifies gravity: with comments on Tsallis
entropy, Eur. Phys. J. C \textbf{82} (2022), 1066 [arXiv:2207.09271 [gr-qc]].
\bibitem{Roos} M. Roos, Introduction to Cosmology (John Wiley and Sons, UK, 2003).
\bibitem{Hehl} F. W. Hehl, P. Von Der Heyde, G. D. Kerlick and J. M. Nester, General Relativity with Spin and Torsion: Foundations and Prospects, Rev. Mod. Phys. \textbf{48} (1976), 393-416.
\bibitem{Capozziello}Francesco Bajardi, Daniel Blixt and Salvatore Capozziello, Hamilton equations in teleparallel gravity and in new general relativity, Phys. Rev. D \textbf{111} (2025), 084012 [arXiv:2412.20592 [gr-qc]].
\bibitem{Weitzen}Weitzenb{\"{o}}ck R., Invarianten Theorie, (Nordhoff, Groningen, 1923).
\bibitem{Maluf} J. W. Maluf, Hamiltonian formulation of the teleparallel description of general relativity, J. Math. Phys.  \textbf{35}, (1994) 335	[arXiv:gr-qc/0002059].
\bibitem{Bengochea}G. R. Bengochea and R. Ferraro, Dark torsion as the cosmic speed-up, Phys. Rev. D \textbf{79} (2009), 124019 [arXiv:0812.1205 [astro-ph]].
\bibitem{Wu}P. Wu and H. W. Yu, f(T) models with phantom divide line crossing, Eur. Phys. J. C \textbf{71} (2011), 1552 [arXiv:1008.3669 [gr-qc]].
\bibitem{Saridakis} S. Nesseris, S. Basilakos, E. N. Saridakis and L. Perivolaropoulos, Viable f(T) models are practically indistinguishable from LCDM, Phys. Rev. D \textbf{88} (2013) [arXiv: 1308.6142[astro-ph.CO]].
\end{thebibliography}
\end{document}